\documentclass[conference]{IEEEtran}
\usepackage{cite}
\usepackage{amsmath,amssymb,amsfonts}
\usepackage{algorithmic}
\usepackage{graphicx}
\graphicspath{{figs/}}
\usepackage{textcomp}
\usepackage{xcolor}
\usepackage{comment}
\usepackage{array}
\usepackage{enumitem}
\usepackage{textcase}
\usepackage[tablename=Table]{caption}
\DeclareCaptionTextFormat{up}{\MakeTextLowercase{#1}}
\usepackage{multirow}

\newcolumntype{C}[1]{>{\centering\arraybackslash}p{#1}}
\usepackage{subcaption}
\usepackage{graphicx} 

\def\BibTeX{{\rm B\kern-.05em{\sc i\kern-.025em b}\kern-.08em
		T\kern-.1667em\lower.7ex\hbox{E}\kern-.125emX}}

\usepackage{authblk}
	
\begin{document}
	
	\renewcommand {\thetable}{\arabic{table}}
	\renewcommand{\thefootnote}{\fnsymbol{footnote}}
	\title{Understanding Public Opinion on Using Hydroxychloroquine for COVID-19 Treatment via Social Media\\
	}

\author[1]{Thuy T. Do}
\author[2]{Du Nguyen}
\author[3]{Anh Le}
\author[4]{Anh Nguyen}
\author[4]{Dong Nguyen}
\author[5]{Nga Hoang}
\author[6]{Uyen Le}
\author[6,*]{Tuan Tran}
\affil[1]{Dept. of Computer Science, UMass Boston, MA. USA. (Email: {\tt thuy.do001@umb.edu })} 
\affil[2]{Dept.of Nursing, Metropolitan State University of Denver (Email: {\tt dnguye79@msudenver.edu})} 
\affil[3]{FPT University, Ha Noi, Viet Nam (Email: {\tt anhltse.fpt@gmail.com})}  
\affil[4]{SaolaSoft Inc., Denver, CO, USA (Email: {\tt \{anguyen, dnguyen\}@saolasoft.com})}  %
\affil[5]{University of Colorado Boulder, Boulder, CO, USA (Email: {\tt nga.hoang-1@corolado.edu})} 
\affil[6]{California Northstate University, Elk Grove, CA, USA (Email: {\tt \{uyen.le, tuan.tran\}@cnsu.edu})}  %
\affil[*]{Corresponding author}  %


%
%
%
%
%

	\maketitle
	\begingroup\renewcommand\thefootnote{\textsection}
	\begin{abstract}
		
		Hydroxychloroquine (HCQ) is used to prevent or treat malaria caused by mosquito bites. Recently, the drug has been suggested to treat COVID-19, but that has not been supported by scientific evidence. The information regarding the drug efficacy has flooded social networks, posting potential threats to the community by perverting their perceptions of the drug efficacy. This paper studies the reactions of social network users on the recommendation of using HCQ for COVID-19 treatment by analyzing the reaction patterns and sentiment of the tweets. We collected 164,016 tweets from February to December 2020 and used a text mining approach to identify social reaction patterns and opinion change over time. Our descriptive analysis identified an irregularity of the users' reaction patterns associated tightly with the social and news feeds on the development of HCQ and COVID-19 treatment. The study linked the tweets and Google search frequencies to reveal the viewpoints of local communities on the use of HCQ for COVID-19 treatment across different states. Further, our tweet sentiment analysis reveals that public opinion changed significantly over time regarding the recommendation of using HCQ for COVID-19 treatment. The data showed that high support in the early dates but it significantly declined in October. Finally, using the manual classification of 4,850 tweets by humans as our benchmark, our sentiment analysis showed that the Google Cloud Natural Language algorithm outperformed the Valence Aware Dictionary and sEntiment Reasoner in classifying tweets, especially in the sarcastic tweet group.

	\end{abstract}
	
	\vspace{-0.1cm}
	\section{Introduction}
	
	
	Hydroxychloroquine (HCQ) is known as a medication to treat and prevent malaria. It is also used for the treatment of rheumatoid arthritis, lupus, and porphyria cutanea tarda \cite{b4}. 
	During the spreading of COVID-19 viruses in 2020, there was some discussion on the effectiveness of using HCQ in treating COVID-19 in some cases \cite{b4}. However, there were no clinical trials with a sufficiently large cohort to provide concrete evidence on the effectiveness of the drug on COVID-19 treatment.
	
	Despite lacking scientific evidence on the efficacy of the drug, \textit{using \underline{H}CQ for \underline{C}OVID-19 treatment} (or \textit{H4C} for short) quickly became a hot topic dominating social media and news. All clinical trials conducted during 2020 found that the drug was ineffective and might cause severe side effects for COVID-19 patients \cite{b4}. This misleading information may put pressure on healthcare systems and society. On one hand, high demand for the drug may be escalated, making it unavailable for prescribed patients. Moreover, COVID-19 patients use the drug for treatment may result in severe side effects that could overload the healthcare systems. Understanding the viewpoints of the community on H4C would help the public health policymakers to develop preventive measures and policy to guide and provide safety to society.
	
	\par
	Current tools such as web-based questionnaire surveys or phone interviews to collect the data from the community are time-consuming, labor-intensive, and costly. Moreover, the long delays of data gathering can make the time-critical decisions suffered. 
	It is important to develop an effective method to collect data and extract the opinions of society. In this study, we proposed to utilize social media to accomplish this goal.
	\par
	By October 2020, Twitter has more than 47 million accounts from the US with 56 percent males and 44 percent females (updated on 10/10/2020) \cite{b15}. Real-time monitoring of public health based on data from social media is promising. In addition, thanks to the availability of APIs and services, collecting data from social media platforms is straightforward. In this study, we analyzed the tweets posted on Twitter to understand the opinions of social media users, and society in general, on the use of HCQ for COVID-19 treatment. We conducted both descriptive analysis and sentiment analysis to reveal the hidden reaction patterns and the shifting of their perceptions on H4C over time. We linked the tweets and Google keyword search frequencies to shed light on the hidden information of the users' opinions on the topic in the space domain. We also evaluated and compared the performance of the state-of-the-art sentiment analysis tools including Google Cloud Natural Language API (GCNL) and Valence Aware Dictionary and Sentiment Reasoner Python library (VADER) on the tweets as well.
	
	\par
	There is some existing work studying online discussions on hydroxychloroquine for COVID-19 treatment. The authors in \cite{b2} calculated the number of tweets mentioning this drug per day from Feb 28 to May 22, 2020, on Twitter to reveal the patterns. They also computed the average sentiment per day to understand the opinions of users on the topic. They found that peaks of reactions on HCQ posts appeared after the days' Trump promoted HCQ on social media. 
	In another study \cite{b6}, the authors analyzed Twitter discussions and emotions using a machine learning approach. In this study, a tweet was classified into one of the eight classes of emotions and one of the thirteen topics to understand the users' opinions. Data showed that ``anticipation" was the most dominant theme while ``surprise" is the least across all 13 topics. Furthermore, the authors in \cite{b7} studied the impact of Trump's promotion of HCQ for COVID-19 patients by analyzing social media content.
	It's reported that the frequencies substantially increased after Trump's discussions about HCQ. However, all of these studies limited their findings in a very short period (\cite{b7} has only 2 months) and that may not be sufficient to reveal the changing of the opinions associated with the development of the pandemic.
	
	\par
	Our work expands the existing frameworks by collecting a more complete dataset spanning in much longer time duration (10 months). In addition, we conducted both descriptive analysis and sentiment analysis of the tweets to understand the opinions of users over time. To the best of our knowledge, we are one of the first studies to link tweets, Google keyword search frequencies, and data from the Centers for Disease Control and Prevention (CDC) to reveal the users' reaction patterns on H4C. Finally, we conducted a manual classification of 4,850 tweet sentiments to evaluate and compare the existing state-of-the-art sentiment analysis tools including GCNL and VADER. In summary, our contributions in this study include:
	\begin{enumerate}
		\item {\bf More Complete Dataset}: We collected 164,016 HCQ related tweets from February to December of 2020 in our study. The collected data provides a more complete picture of society's perspectives on the use of HCQ for COVID-19 treatment. This is one of the most complete datasets on the topic that has been collected so far.
		\item {\bf Identifying Reactions Patterns in both Time and Space Domains}: We conducted both descriptive and sentiment analysis in both time and space domains to reveal the reaction patterns of both online and geographically local communities on H4C. 
		\item {\bf Linking Multiple Data Sources to Reveal Hidden Reaction Patterns}: We also linked data from Twitter, Google, and CDC to identify reaction patterns and the relationship between ``listening" (reactions on Twitter) and ``doing" (search queries on Google) and ``did" (purchased drug, CDC reports). 
		\item {\bf Conducting Manual Classifier}: In this study, we manually classified 4,850 tweets associated with important events of the HCQ and COVID-19 developments to evaluate and compare the existing sentiment analysis tools. To our best knowledge, this is one of the largest US-based users datasets of tweets regarding COVID-19 and HCQ. We plan to share this dataset with the research community upon completion of this project.
	\end{enumerate}
	
	The remainder of the paper is as follows. In Section II, we present our system architecture and data processing workflow. In Section III, we describe our data analysis methodology. We then describe the research results and discussions in Section IV. Finally, we provide some concluding remarks and future directions in Section V.
	\begin{figure*}[th!]
		\centering 
		\includegraphics[height = 3.8 cm, width=1.5\columnwidth]{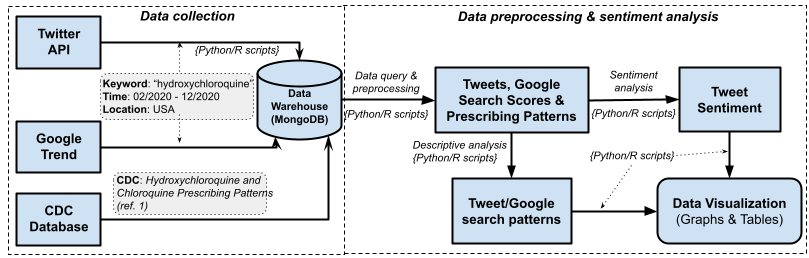} 
		\caption{System architecture for data collection and analysis.}
		\label{fig:system_architecture} 
		\vspace{-0.15in}
	\end{figure*}
	
	\begin{figure}[th!]
		\centering 
		\includegraphics[width=1\columnwidth]{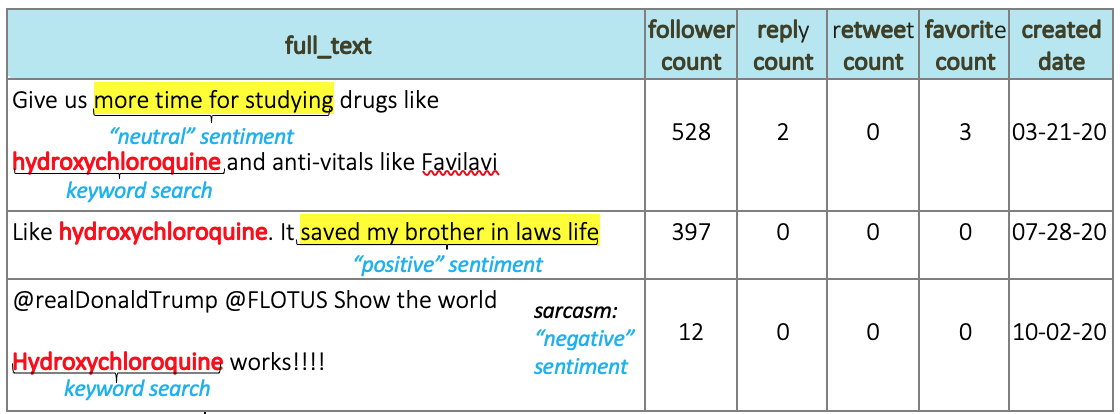} 
		\caption{Samples of tweets.}
		\label{fig:tweet_sample} 
		\vspace{-0.2in}
	\end{figure}
	
	\section{System Architecture and Data Processing Workflow}
	
	Our system architecture and data processing workflow is illustrated in Fig. \ref{fig:system_architecture}. The system consists of two main components:
	
	\begin{itemize}
		\item {\bf Data collection:} The first component of the system is ``Data collection". In this process, we used different techniques to collect data from Twitter, Google Trend, and the Centers for Disease Control and Prevention (CDC) database. 
		\begin{itemize}
			\item {\it Twitter API:} To collect related tweets on Twitter \cite{b15}, we developed a Python script that connected to the Twitter API to search for related tweets. We used ``hydroxychloroquine" as the keyword, duration from February 2020 to December 2020 as time period, and the U.S. as the location for our query. The retrieved objects of the query were in the JSON format that includes metadata of the tweets consisting of the text of tweets, tweets' time stamp, reactions (e.g., ``love", ``favorite"), etc. 
			\item {\it Google Trend:} We also queried Google Trend platform \cite{b18} to collect data of keyword search regarding hydroxychloroquine as well. We used the same keyword, time duration, and location as specified in the query used on Twitter. The retrieved data consists of search scores and time stamps of the search volumes in different states. We should emphasize that the search scores were normalized, ranging from 0 (no search queries) to 100 (maximum search queries).
			\item {\it CDC Prescribing Patterns:} We also collected prescribing patterns of hydroxychloroquine and chloroquine from the database of CDC between January and June of 2019 and 2020 \cite{b8}. We used prescribing data of both drugs as they are clinical equivalence in treatment.
			
		\end{itemize}
		\noindent All the collected data from the sources are stored locally in our data warehouse using MongoDB. 
		
		\item {\bf Data preprocessing and sentiment analysis:} The second component of our system architecture is the ``Data pre-processing and sentiment analysis''.  
		
		\begin{itemize}
			\item{\bf Data query and pre-processing:} Text mining on social media is challenging due to unstructured and noisy data \cite{b21}. Thus, before analyzing the data for patterns and text sentiment, we filtered out noise using database query and data pre-processing.  
			\begin{itemize}
				\item {\it MongoDB query:} We queried our local database to extract related tweets and converted the retrieved tweets into a table format where each row represents a tweet and each column represents an attribute of the tweet (e.g., text, timestamp, etc.). 
				A sample of the tweets is illustrated in Fig. \ref{fig:tweet_sample} with highlighted search keyword and sentiment words.
				\item {\it Data pre-processing:} Next, we developed R and Python scripts to pre-process the tweets before feeding the data to the algorithms for performing descriptive analysis and sentiment analysis. The following procedures are performed in our pre-processing step:
				\begin{itemize}
					\item{\it Remove non-English tweets}: In the first step, we remove all tweets having the keyword but written in different languages. This is to ensure the count of tweet frequencies and their sentiments consistent. We use the Python package ``Enchant" \cite{b16} to detect and delete words of tweets not in English.
					\item {\it Remove duplication, punctuation, URLs, HTML tags and entities (e.g., \&amp;)}: In this step, we used a search and lookup script written in Python to remove all punctuations, URLs, HTML tags, and entities (e.g., \&amp;) which do not contribute to the sentiment of the tweets. In addition, we compared the ID and time stamp to remove duplicated tweets in this step as well. This is to prevent spurious sentiment scores due to the duplication of tweets.
					
					\item {\it Word Lemmatisation and Stop-word Removal:} In addition, we converted words from different forms to their root forms. For instance, \textit{``happier/Happier"}, \textit{``happiest/Happiest"} and \textit{``happily/Happily"} are converted to its original form ``happy". In other words, word lemmatization is a text normalization that reduces the redundant dimensionality of the text. This step is important to ensure the accuracy of our sentiment analysis in the next step. We also performed stop-word removal at this step as well. Stop-words are commonly used words but not contributing to the sentiment of a sentence (e.g, \textit{``the"}, \textit{``a/an"}, \textit{``of"}, etc.). We adopted the well-known Natural Language Toolkit (NLTK) \cite{b17} for text lemmatization and stop-word removal.
					
					\item {\it Emojis and Emoticons Conversion}: Finally, we observed that the collected data consisted of several tweets with emojis and emoticons \cite{b9} embedded in the text. Without pre-processing these emojis and emoticons, the sentiment analysis may not be accurate or might be interpreted in opposite meaning. For example, tweet \textit{``just used hydroxychloroquine, feeling :-)``}. Removing the smiley face ``:)'' emoticon, the sentiment of the tweet should be \textit{``neutral"}. On the other hand, if we converted it to \textit{``happy''}, the actual meaning of the emoticon here, the sentiment of the tweet changed to \textit{``positive"}. We developed a Python script that used a lookup table described in \cite{b10} to convert all emojis and emoticons for all tweets.
				\end{itemize}
			\end{itemize}
			\item {\bf Descriptive Analysis, Sentiment Analysis and Data Visualization:}  The pre-processed data is fed into two algorithms for descriptive analysis and sentiment analysis (more detail is described in the next section). The outputs of these two algorithms are gathered and displayed in the forms of graphs and tables described in Section \ref{sec:results}. 
		\end{itemize}	
	\end{itemize}
	
	\section{Data Analysis Methodology}  
	
	Our data analysis methodology consists of two parts. 
	\begin{itemize}
		\item {\bf Descriptive Analysis}: First, we perform data descriptive analysis by visualizing the pre-processed data to observe the trend and patterns of the tweets and Google keyword search over time. We also used the prescription orders collected from the CDC to observe the purchase patterns of the drug.	
		\item {\bf Sentiment Analysis}: Second, we perform sentiment analysis of the tweets to reveal the opinion of the users on H4C. We should emphasize that extracting sentiment of noisy tweets is challenging due to the short texts and embedded emojis and emoticons \cite{b19}. 
		To quantify the users' opinion on support or against the use of ``hydroxychloroquine'' for COVID-19 treatment, tweets are classified into three categories: Positive (\textit{Pos}) (a supporting opinion), Negative (\textit{Neg}) (an opposition), and Neutral (\textit{Neu}) (neither support nor against, general statement of using the drug). In this analysis, our goal is two folds: (1) revealing the opinion of users on H4C, and (2) comparing the sentiment classification performance of existing well-known sentiment analysis tools, i.e., VADER and Google Cloud Natural Language API. 
		
		\begin{itemize}
			\item {\it Manual Sentiment Classifier (MSC)}: In MSC, five undergrad students spent total 50 hours to read and classify tweets in different categories ({\it ``Pos''}, {\it ``Neg''}, or {\it ``Neu''}). Due to the large size of the collected dataset (164,016 tweets), we proportionally randomly selected 4,850 tweets posted in five important dates associated with the highest numbers of tweets sent on Twitter, including March 21, April 06, May 19, July 28, and October 02 (see Fig. \ref{fig:1_num_tweets_favorites_ggscore_vs_time} and Fig. \ref{fig:important_dates}  for detail). We adopted guidelines for classification task from \cite{b4}. 
			To ensure the consistency of tweet classification across all the students, we did pre-training on interpreting the meaning of tweets, especially tweets with sarcastic meaning. For example, the tweet \textit{``Just give Trump hydroxychloroquine and send him on his way. You know the miracle cure"} is identified as sarcasm because it was created on the day Trump was tested positive with COVID-19, October 02. As a result, this tweet should be interpreted as \textit{``Neg''} as the user implied that HCQ did not work on treating COVID-19. In addition, the second round of cross-checking and group discussion were performed for tweets where the first student could not determine their meanings. The manually classified dataset is published on Github.\footnote{https://github.com/thuydt02/HCQ\_Tweet\_Dataset}
			\item {\it Valence Aware Dictionary and sEntiment Reasoner (VADER)}: The second technique used to classify tweets was the VADER Python library \cite{b12}. VADER is a lexicon and rule-based sentiment analysis program that was particularly attuned to analyzing social media text. It implemented 7,500 lexical features with validated valence scores on the scale from -4 (extremely negative ) to 4 (extremely positive), with the midpoint 0 as neutral. As an example, the valence score for ``great" is $3.1$, ``summer" is $0$, and ``horrible" is $-2.5$. VADER calculates a sentiment metric consisting of four elements: Positive, Negative, Neutral, and Compound. The first three elements represent the proportion of the text that falls into those categories, ranging from 0 to 1, inclusively. The final compound score (\textit{ComS}) is the sum of all of the lexicon ratings and normalized to a range between -1 (most negative) and 1 (most positive). Mathematically, the \textit{ComS} is computed by\\
			\begin{equation}
				ComS = \frac{x}{\sqrt{x^2 + \alpha}},
			\end{equation}
			where  $x= \sum_{i=1}^{n}s(w_i)$, $s(w_i)$ is the valence score of $i_{th}$ word in the text, $n$ is the total number of words in the text; and $\alpha$ is the normalization constant (default value is $15$). We notice that due to the noise of the media content, we adopted the thresholds proposed in \cite{b10}\cite{b11} where a tweet is classified as \textit {``Pos'', ``Neu'', ``Neg''} if its \textit{ComS $\geq$ 0.05, 0.05$>$ ComS $>$ -0.05, ComS $\leq$-0.05}, respectively. 
			\item {\it Google Cloud Natural Language API (GCNL)}: Our third technique is the well-known advanced machine learning Google Cloud Natural Language API \cite{b13} which has a pre-trained model for sentiment analysis, called ``{\it analyzeSentiment}". It identifies the prevailing emotional opinion within the text, especially to determine a writer's attitude as positive, negative, or neutral. The sentiment metric by {\it analyzeSentiment} has two factors: score (\textit{GScore}) and magnitude (\textit{GMag}). Similarly the \textit{ComS} of VADER, \textit{GScore} is calculated to determine the sentiment polarity of the text with its range from -1.0 (most negative) to 1.0 (most positive). On the other hand, the \textit{GMag} is used to represent the overall strength of emotion of the text, ranging from 0.0 to \textit{+inf}. Unlike score, \textit{GMag} is not normalized; each expression of emotion within the text contributes to the text's magnitude. Thus, longer text blocks will likely have greater magnitudes. We use the same threshold defined in VADER to classify the classes of the tweets. We expected that GCNL with an advanced machine learning algorithm should perform well in understanding the tweets to identify the users' opinions on using HCQ for COVID-19 treatment.
			
		\end{itemize}
	\end{itemize}
	
	\begin{figure}[tb]
		\centering 
		\includegraphics[width=1\columnwidth]{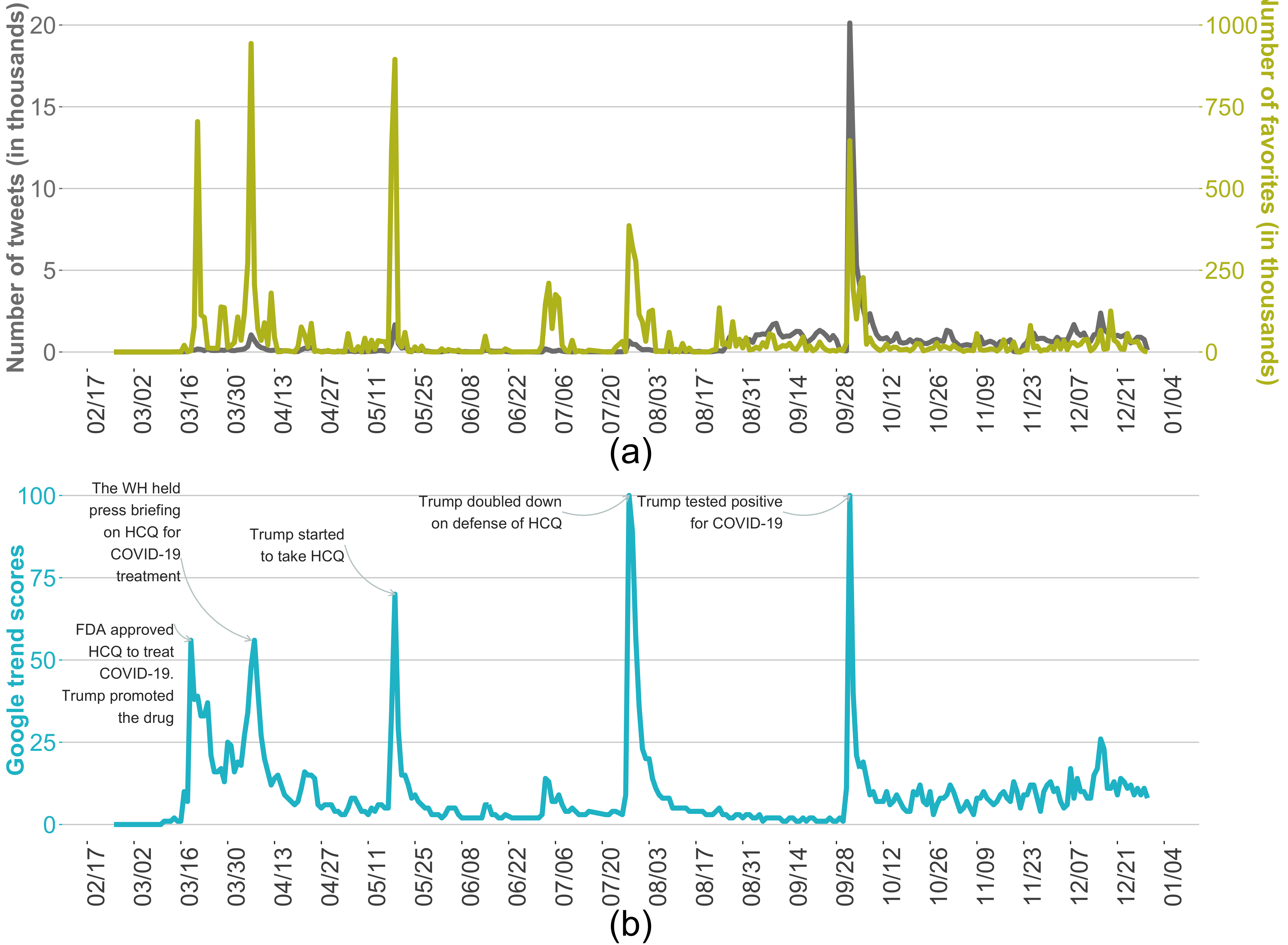} 
		\caption{Abnormal online users' reaction patterns in the time domain. (a) Number of tweets, favorites on Twitter, (b) Google keyword search score.}
		\label{fig:1_num_tweets_favorites_ggscore_vs_time} 
		\vspace{-0.15in}
	\end{figure}
	
	\begin{figure}[tb]
		\centering 
		\includegraphics[height = 4 cm, width=.9\columnwidth]{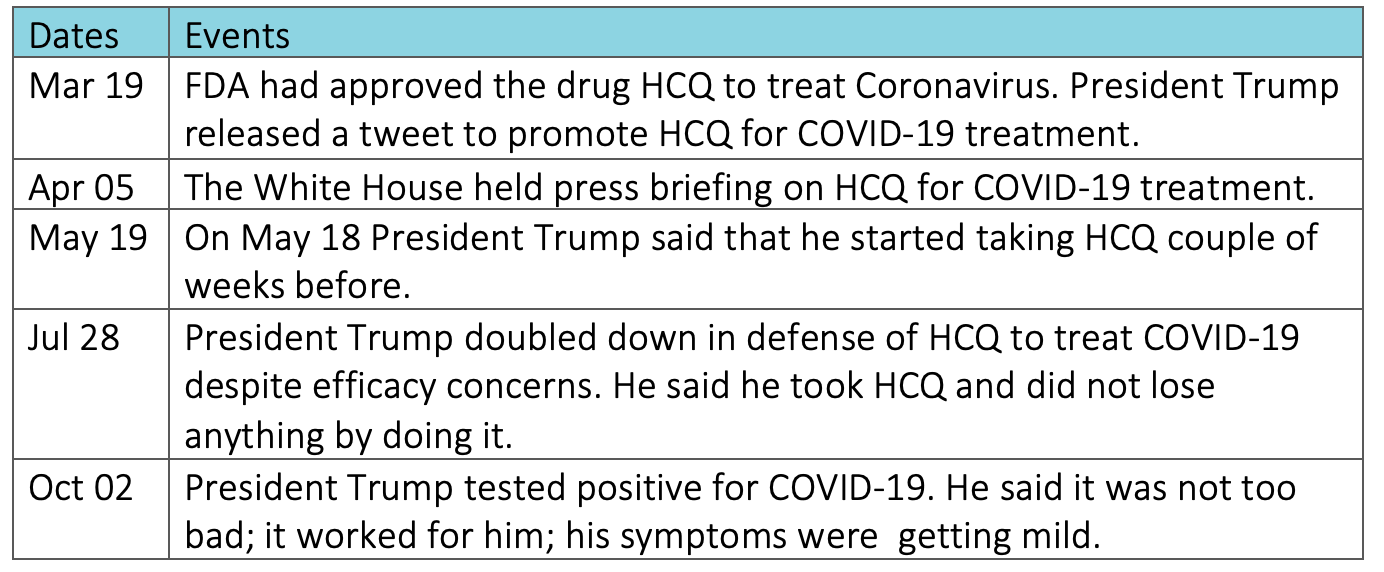} 
		\caption{Social and news events related to using HCQ for COVID-19 treatment.}
		\label{fig:important_dates} 
		\vspace{-0.2in}
	\end{figure}
	\vspace{-0.1in}
	\section{Results and Discussion}
	\label{sec:results}
	
	\subsection{Descriptive Analysis}
	We first identify the trend of the reactions of online users via tweets and Google keyword search frequencies. We hope that the reaction patterns may shed some light on how social media users react in response to the information feeds regarding using HCQ for treating COVID-19.
	
	\subsubsection{Unexpected Patterns of Users' Reactions} 
	We first plot the reactions of users on Twitter and Google search platform in the time domain in Fig. \ref {fig:1_num_tweets_favorites_ggscore_vs_time}(a) and (b), respectively. Our data shows unexpected patterns on both of the platforms with reaction spikes that occurred on only some specific dates and instantly diminished right after these peaks. The observed pattern is unexpected and it revealed interesting patterns in how online users react to news and social media feeds. Intuitively, we expected the reactions to maintain at some levels for a longer duration. We also would like to emphasize that the reaction patterns perfectly align in the time domain between the Twitter and Google platforms. That sheds light on the hidden link between "listening" on social media (feeds on Twitter and news) and "taking actions" on searching for information (on Google).

	\subsubsection{Revealing Emerging Society Interests}
	From the observed data patterns from the tweet and Google keyword search frequencies, we further explored to understand the "spike" reaction patterns of the online users. We identified five social and news feeds listed in Fig. \ref{fig:important_dates} that helped to explain the patterns of the social media users' reactions. Furthermore, the data showed that October 02 had the highest number of reactions with 20,124 tweets, about 12 times higher than other dates which had about 1,663 or fewer posts on average. This abnormality reveals the current interest of the society - COVID-19 treatment and controversial statements by President Trump who was announced being positive to COVID-19 on that day.
	
	\subsubsection{Online Users' Reaction in the Space Domain}
	We are also interested in how the online users' reactions to the information and news feed of using HCQ for treating COVID-19 in the space domain. The Google Trend keyword search frequencies across different states are illustrated in Fig. \ref{fig:2_gg_trend_vs_states}. As we observed, the midwest and mountain states of the U.S had the most keyword search frequencies. Particularly, South Dakota and Montana were the two states with the highest keyword search frequencies.
	The data can be interpreted as there were higher degrees of interest from the communities geographically located in these areas regarding H4C. It also sheds light on how the local public health policy was conducted (e.g., South Dakota was one of the first states to test HCQ on COVID-19 treatment) and the political viewpoints of the communities in respect to their parties regarding their opinions on using HCQ for COVID-19 treatment. This is one of the first studies that reveal this hidden link using social media data.   
	\begin{figure}[tb]
		\centering
		\includegraphics[width=0.9\columnwidth]{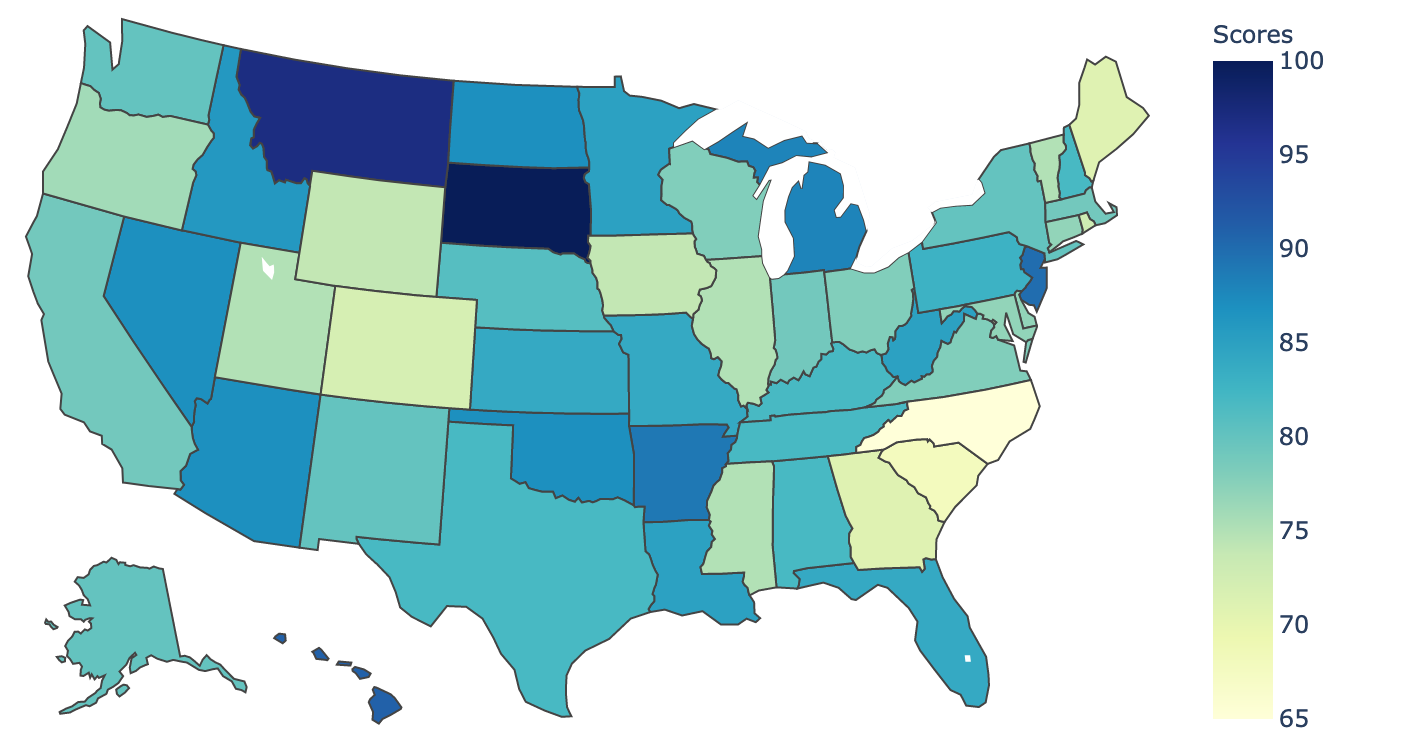} 
		\caption{Google Trend keyword search scores by states.} 
		\label{fig:2_gg_trend_vs_states} 
	\end{figure}
	
	\subsubsection{Linking Social Media Reactions to the Drug Purchase Actions}
	Our Twitter and Goodle data showed hidden patterns of how social media users reacted to HCQ. However, it's not clear if the public actually purchased the drugs for COVID-19 treatment. To answer this question, we collected data of the prescriptions of hydroxychloroquine/chloroquine from January to June in 2019 and 2020 from the Centers of Disease Control and Prevention (CDC) \cite{b8}\footnote{Prescription data from June to Dec were not available so it was absent from our plot.}. We observed a significant increase (i.e., about 10 times in March 2020 and 6 times in April) in the number of prescriptions for the drug in 2020 compared to that of 2019. 
	This data shows clear evidence that the public took action on purchasing the drug for treatment consideration. Interestingly, we also observed some decline in the number of routine and non-routine prescriptions in the following months. This can be explained by the new evidence and studies showing that the medication was not effective in COVID-19 treatment\cite{b8}.
	
	\begin{figure}[tb]
		\centering 
		\includegraphics[height= 4.2 cm, width=0.85\columnwidth]{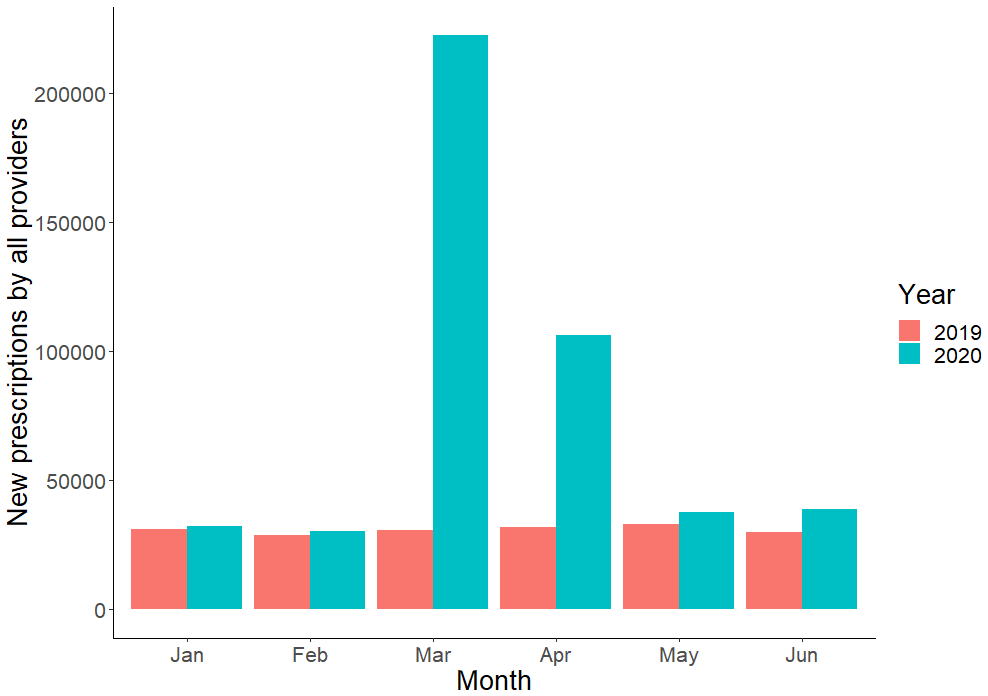} 
		\caption{New prescriptions by all providers (routine, primary care and nonroutine.}
		\label{fig:3_prescriptions_bar}
		\vspace{-0.2in}
	\end{figure}
	
	\subsection{Sentiment Analysis}
	The descriptive analysis in the previous subsection revealed some big picture of how social media users reacted to news feeds on using HCQ for COVID-19 treatment. However, the descriptive analysis did not provide sufficient information to answer the following questions: \textit{ (1) What was the overall opinion of social media users on using HCQ for COVID-19 treatment? (2) How did the degree of support/against change over time? (3) Do the existing sentiment analysis tools work well on the noisy dataset?} In this subsection we performed sentiment analysis of the tweets to shed light on the hidden information in the tweets to find the answers for these above questions. 
	
	\subsubsection{Extracting Opinions via Word Frequencies}
	We first investigate the word frequencies of the tweets to reveal the crowd opinion as a whole. The wordcloud of the tweets is plotted in Fig. \ref{fig:wordcloudposneg}. As we observed in Fig. \ref{fig:wordcloudposneg}(a), ``treatment",  ``taking", ``cure" were standing out as the most frequently used words in the tweets. This can be interpreted as, in general, social media users supported the recommendation of using HCQ for COVID-19 treatment. Additionally, in Fig. \ref{fig:wordcloudposneg}(b), we plotted the wordcloud associated with the sentiment of the words. Particularly, we used the ``bing" lexicon \cite{b20} to classify the words into the positive and negative classes. As a result, the neutral words are excluded from the data. Interestingly, we observe that positive words (the lower half in green color) dominated the negative words, and ``cure" is the most frequently used word in the positive sentiment class.

	\begin{figure}
		\centering
		\includegraphics[height = 4.5 cm, width=0.7\linewidth]{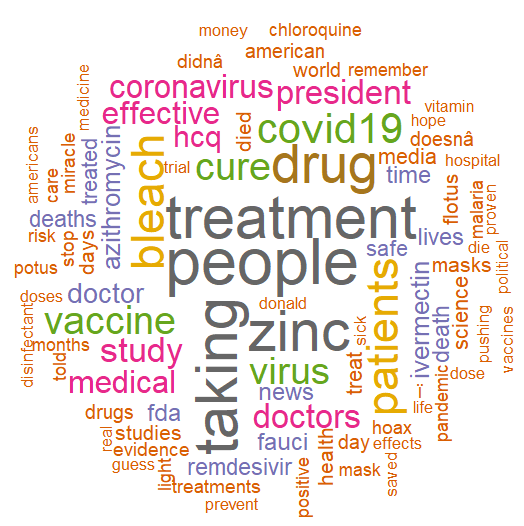} \\
		(a) \\
		\includegraphics[height = 4.5 cm, width=0.7\linewidth]{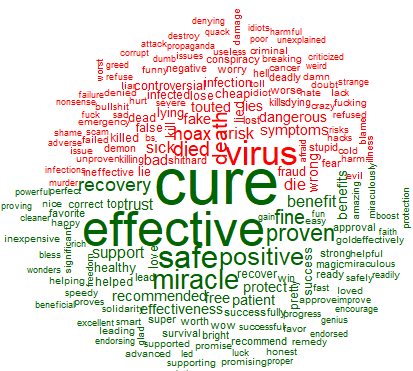} \\
		(b)
		\caption{Wordcloud of the tweets. (a) Most frequently used words, (b) Positive vs. negative words using "Bing" lexicon.}
		\label{fig:wordcloudposneg}
		\vspace{-0.2in}
	\end{figure}
	
	\subsubsection{Quantifying the Change of Opinions}
	Next, we investigate how the opinions of the social media users shifted over time by quantifying the shift of positive and negative sentiment tweets over time. We focus on five important events that we identified in the Descriptive Analysis that associated with the most numbers of reactions from the users. To ensure the conclusion drawn from this step reliable, we only considered the 4,850 tweets randomly selected from these dates using a manual sentiment classifier (MSC). Fig. \ref{fig:human_ratio_pos_to_neg_vs_time} shows the ratio of positive sentiment tweets to negative sentiment tweets over time. Generally speaking, we see the opinion of users shifted from less support in March (the ratio is less than 1) to more support in April and July (the ratio is greater than 1). We also observed that the negative opinion dominated in May and October. This might be due to the tweets sent out on these dates more related to an individual, Donald J. Trump, than the use of HCQ for COVID-19 treatment. It's an interesting data pattern for further investigation in the future.
	
	\begin{figure}[tb]
		\centering
		\includegraphics[height = 4.5cm, width=0.85\columnwidth]{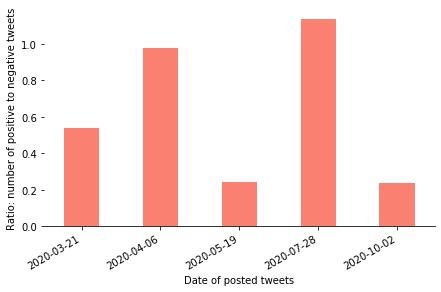} 
		\caption{Shifting of social media users' perception on using HCQ for COVID-19 treatment in the time domain.} 
		\label{fig:human_ratio_pos_to_neg_vs_time} 
		\vspace{-0.2in}
	\end{figure}
	
	\subsubsection{Sentiment Classification Comparison} Finally, we evaluate the performance of GCNL and VADER algorithms which are efficient in processing large datasets by comparing their sentiment classification accuracy with the MSC. Particularly, we used 4,850 randomly selected tweets from the five events with the most number of reactions for our comparison. The sentiment classifications of GCNL and VADER are illustrated in Fig. \ref{fig:3_human_vader_gg_sentiment}. Here the performance of MSC  classified by humans is set as the benchmark with 100\% accuracy. As we can see, GCNL and VADER do not perform well in this dataset. That is because they may not recognize sarcastic tweets. In addition, GCNL performs slightly better with an average accuracy of 42.5\% compared to 38.7\% of VADER. We also would like to emphasize that the event of October 2nd had many sarcastic tweets and GCNL significantly outperformed VADER by 9\% thanks to its advanced machine learning algorithm in natural language processing. To the best of our limited knowledge, this is one of the first sentiment analysis studies comparing existing tools against human classification on a large tweet dataset.
	
	
	\begin{figure}[tb]
		\centering
		\includegraphics[height = 5 cm, width=0.85\columnwidth]{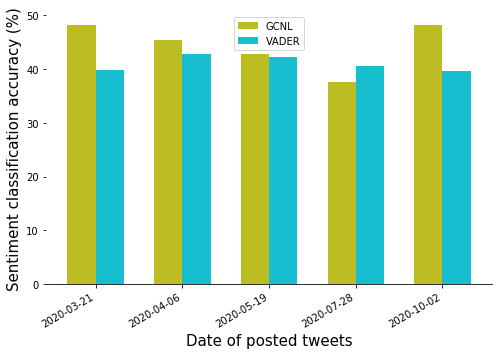} 
		
		\caption{Performance comparison of sentiment classification of GCNL and VADER. }
		\label{fig:3_human_vader_gg_sentiment} 
		\vspace{-0.2in}
	\end{figure}
	

	
	\section{Conclusion}
	Mining text on social media to understand the online users' opinions is challenging. In our study, we collected 164, 016 tweets posted in 2020 with the ``hydroxychloroquine'' (HCQ) keyword on Twitter to extract the opinions of online social users on the recommendation of using HCQ for COVID-19 treatment. Our descriptive analysis identified an irregularity of users' reaction patterns that are tightly associated with the related social media feeds and news on the development of HCQ and COVID-19 treatment. The study linked the tweet and Google keyword search frequencies to reveal the viewpoints of communities on H4C located in different geographical locations across different states. In addition, we analyzed the sentiment of the tweets to understand the public opinion on the recommendation of using HCQ and how it changed over time. The data shows that high support in the early dates but it declined over time.  
	Finally, our sentiment performance comparison showed that GCNL outperformed VADER in classifying tweets, especially in the sarcastic tweet group. We will further utilize the social links and friend counts of the users to characterize how misinformation spreads out in the social media network in our future study.
	
	\vspace{12pt}
	\color{red}
\end{document}